\documentclass[nonacm, manuscript]{acmart}

\AtBeginDocument{%
  }

\usepackage{graphicx}
\usepackage{longtable}
\usepackage{caption}       % for caption formatting
\usepackage{subcaption}
\usepackage{array}
\usepackage{booktabs}      % for better looking tables
\usepackage{multirow} 
\usepackage{tabu}      
\usepackage{tabularx}

\begin{document}

\title{A Relational (Re)Turn: Revisit Interactive Art through Interaction and Aesthetics}

\author{Aven-Le ZHOU}
\email{aven.le.zhou@gmail.com}
\orcid{0000-0002-8726-6797}
\affiliation{%
  \institution{The Hong Kong University of Science and Technology (Guangzhou)}
  \streetaddress{No.1 Du Xue Rd, Nansha District}
  \city{Guangzhou}
  \state{Guangdong}
  \country{P.R.China}
  % \postcode{43017-6221}
}

\renewcommand{\shortauthors}{Aven-Le Zhou.}

\begin{abstract}
This paper revisits the concept of interaction in interactive art, tracing its evolution from sociocultural origins to its narrowing within human-computer paradigms. It critiques this reduction and proposes a relational (re)turn through reclaiming interaction as intersubjective and relational. Through a synthesis of aesthetic theories and case studies from Ars Electronica, the paper introduces Techno Relational Aesthetics, a new conceptual lens that emphasizes technologically mediated relationality. This approach expands interactive art beyond audience–artwork interaction and opens the possibility to broader relational practices.
\end{abstract}


\maketitle

\section{Introduction}

The \textit{``Dictionary of Philosophy and Psychology''} defines ``interaction'' as ``the relation between two or more relatively independent things or systems of change which advance, hinder, limit, or otherwise affect one another \cite[p.561]{baldwin_dictionary_1901},'' which refers to both the body-mind relationship and the interaction of objects in and with the environment \cite[p.16]{kwastek_interactivity_2008}. Kwastek further refines this term in the context of art, describing interaction as a ``mutual or reciprocal action or influence'' and emphasizing ``reciprocity'' as key to understanding the shift toward participatory form \cite[p.16]{kwastek_interactivity_2008}. This aesthetic of interaction transforms the viewer from passive observer to active participant, introducing a feedback loop wherein audience behavior influences and shapes the form and behavior of the artwork itself \cite{kwastek_aesthetics_2013}. 

This transformation is not merely formal; it signals a reconfiguration of the audience-artwork relationship through technological mediation. Early interactive artworks often emerged as showcases of experimental technologies, introducing novel techno-experiences through sensors, actuators, real-time computation, or networked environments. The fascination lay in the technological medium's potential to engage the body as an interface, or what somaesthetics frames as the lived, felt experience of the body, detailed through various modalities of somatic perception. Yet these early explorations, while foundational and still compelling in many cases, now risk becoming reductive and increasingly reveal conceptual and technological limitations, particularly when artistic reflection centers narrowly on interactivity and confines body to the individual proper as the dominant framework for understanding the interactive experiences.

This study adopts a critical and interdisciplinary approach, drawing from media art theory and history, philosophical aesthetics, and practice-based analysis. As an artist-scholar working at the intersection of interactive art and theory, I approach the notion of interaction through both conceptual reflection and practical engagement with artworks, tracing trajectories in aesthetics and interaction, while interpreting selected artworks from Ars Electronica as concrete sites of the emerging relational pattern. This dual approach allows me to interrogate how interaction is conceptualized, mediated, and experienced across different views. Through the conceptual and empirical engagements with the evolving discourse of interaction and aesthetics, this paper revisits interactive art and proposes a relational (re)turn. 

In the following sections, I first trace the shifting meanings of interaction from its sociocultural origins to its narrowing within cybernetics and human-computer interaction. Second, I examine the history of interactive art as a genre, outlining its definitional and crisis debates. Third, I revisit aesthetic frameworks and propose Techno Relational Aesthetics as a synthesis that integrates technological mediation with intersubjective experience. Fourth, I analyze case studies from Ars Electronica that exemplify this techno-relational orientation, focusing on audience–audience engagements and relational interaction. Finally, I propose a relational (re)turn in interactive art to embrace the intersubjective dynamics.

\section{Interaction}
\label{sec: interaction}

Before delving into the aesthetics of interactive art, this section examines the shifting conceptualizations of interaction itself and how it evolved from intersubjective and social roots to its narrowing within art and technological contexts. 

% Sec.~\ref{subsec:shifting_scientific_interests} sources the roots of this term, adopted by scientific fields as heterogeneous as social psychology, physiology, sociology, cybernetics, and computer science, i.e., human-computer interaction. Sec.~\ref{subsec:artistic_development} follows Ryszard W. Kluszczynski's ``rhizomatic archipelago'' view of interactive art's origin and reviews the artistic developments of interaction; it contextualizes interactive art within broader contemporary art developments and focuses on how the social and participatory interaction in the arts evolves into the narrowed human-machine interactions. 

\subsection{The Shifting Scientific Interests}
\label{subsec:shifting_scientific_interests}
This section overviews the evolving concept of interaction from the early 1960s. It had developed from an idea of reciprocity in biological, chemical, and physiological processes elaborating theories of social interaction (sociology), to a whole new science trying to establish the concept of feedback processes as a fundamental theorem of life and technology (cybernetics), and eventually to a field of research and development in the computer sciences \cite[p.15]{kwastek_interactivity_2008}. 

\subsubsection{Social Psychology and Sociology}
In the early 20th century, ``interaction'' was applied to social and societal processes as a science of sociology. Georg Simmel was believed first to apply the term ``Wechselwirkung'' (i.e., interaction) to describe interpersonal relationships \cite[p.16]{kwastek_interactivity_2008}. In stimulus-response theory and to explain it physiologically, ``interpersonal interaction'' consists of a complex process of causes and effects of the various sensory organs and muscle groups \cite[p.170]{schmidt_man_1937}. While symbolic interactionists are more interested in actions, stimulus-response theorists principally investigate reactions \cite[p.16]{kwastek_interactivity_2008}. George Mead and Edward Ross discussed ``social interaction'' or the ``interaction of human beings,'' which was later summarized by Herbert Blumer as ``symbolic interactionism'' \cite{blumer_symbolic_1986}. Katja Kwastek points out that ``social interaction is primarily a communicative process in which people share experience, rather than a mere play back and forth of stimulation and response'' from a symbolic interactionist's perspective \cite[p.171]{schmidt_man_1937}. 

\subsubsection{Cybernetic Theory} 
Before the concept of interaction was introduced into the arts, cybernetic theory explored the idea. However, cybernetics, coined by Norbert Wiener in 1947, is less interested in the ``interactions between human beings'' than analogies between the self-organization of living organisms and machines. It studies the control and communication between the animal and the machine, focusing on how digital, mechanical, or biological systems process information and how they react to it. Cybernetics studies feedback mechanisms and information processing across engineering and biology by examining black boxes and derived concepts such as communication and control \cite[p.203]{sommerer_cultural_2011}. This theory of feedback processes goes beyond the stimulus-response theory in distinguishing different types of feedback and eventually leads to self-organization in the individuals \cite[p.16]{kwastek_interactivity_2008}.

Jack Burnham, in 1968, contextualized cybernetic art within art history: ``The spectacle of an artifact adjusting to its environment through a series of visible maneuvers has a certain anthropomorphic fascination'' \cite[p.338]{burnham_beyond_1968}. In ``Cybernetic Spatiodynamic Sculptures'' by Nicholas Schoeffer, in the 1950s,  he used the cybernetic concept to organize the reaction of these works to the environment via sensors. Other artists like James Seawright, Edward Ihnatowicz, and Tony Martin built works that interact with their environment via light and sound sensors and/or installed environments that reacted to the audience, emitting light and/or sound \cite{burnham_beyond_1968}. Yet they did not call their works interactive; instead, they were called cybernetic, responsive, or reactive \cite{kwastek_invention_2008}.

\subsubsection{Human-computer Interaction}
It was not until the beginning of the 1960s that computer science developed to a stage that allowed for real-time interaction between men and computers. In 1960, J. This workstation greatly impacted how we use and interact with computers today and popularized graphical user interfaces (GUIs) altogether'' \cite{licklider_man-computer_1960}. In 1963, computer pioneer Ivan Sutherland developed the ``Sketchpad,'' a graphical interface that enabled manipulating graphics on a display using a light pen. In 1965, Douglas Engelbart created the ``X-Y position indicator for a display system,'' now known as the computer mouse. 

With the principal concept of the graphical user interface developed by Sutherland and Engelbart's mouse replacing the light pen, essential human-computer interface elements were available. From then on, human-computer interaction was established as a highly specialized and interdisciplinary field within computer science \cite{weiser_computer_1999}. Kwastek also recognized it as a heavy influencer in interactive art regarding audience interaction mediated by technology \cite{kwastek_interactivity_2008}. Ernest Edmonds compares HCI in interactive art making to the colors of paint in painting and states that the principal aesthetic value of interactive art lies in the interactive experience \cite{edmonds_art_2010}.

\subsection{The Artistic Developments}
\label{subsec:artistic_development}
Due to the ``nonlinear construction, fluidity, transformation in many planes and directions'' of interactive art history, Ryszard Kluszczynski suggests viewing the development of the field as a ``rhizomatic archipelago'' \cite{kluszczynski_viewer_2013}. Inke Arns also argues that ``the notions and concepts of interaction, participation, and communication are central to twentieth-century art, and in equal measure concern the work, recipient, and artist. Generally, these terms involve an art movement from the closed to the `open' work of art, from the static object to the dynamic process, from contemplative reception to active participation'' \cite{arns_interaction_2004}. 

Art movements such as Dada, Fluxus, Happenings, Kinetic art, OP art, and Cybernetic art have introduced the ideas of audience participation, art as a process, and immateriality of an artwork \cite[p.34]{guljajeva_interaction_2018}. In this section, I review these artistic developments to reveal how the interaction in the arts has been transforming from human relations and participatory experiences to the narrowed human-machine interactions.

\subsubsection{Dada, Flux and Happening}

Marcel Duchamp, the critical figure of Dada, and his remarkable work inspired many artists in the 1960s. He pioneered assigning roles to the machine and the audience to be creative and part of the work. Even though Duchamp did not consider himself a progenitor of ``art and technology,'' he saw the importance of the additional value of the beholder in the creative act \cite{edmonds_art_2010}. Apart from Duchamp, many other artists play crucial roles in expanding existing art forms and mixing them with other disciplines and genres, such as John Cage, Robert Rauschenberg, Merce Cunningham, Nam June Paik, and more \cite{guljajeva_interaction_2018}. Many of these Fluxus and Neo-Dada artists treat instructions as a method for guiding audience participation and give a specific decisive role to the audience or performer(s). 

Technology as a mediator between the artist and the audience was introduced from the 1950s onwards. When technology-mediated participation was introduced, artists started communicating instructions via technological systems. Similarly to Dadaism and Fluxus, interactive art shares the same values regarding bringing art closer to the public rather than keeping it to an expert audience \cite[p.35]{guljajeva_interaction_2018}. Dinkla describes the difference in regarding the artist's role in Happenings versus interactive art: the Happening artist uses the audience as material, whereas in interactive art, the system is open for discovery, and the artist trades their freedom to the audience and to the machine \cite[p.288]{dinkla_participation_1996}. 

\subsubsection{Kinetic art, OP art, and Cybernetic art}

Kinetic and OP art are vital in further developing interactive art since they have addressed the spectator as an active element. ``Everything that would later characterize computer art and the interactive virtual environment was there already, albeit in purely analog or mechanical form'' \cite[p.38]{weibel_it_2007}. Peter Weibel values the most Kinetic and OP art influences in the path of the interactive art establishment \cite[p.35]{guljajeva_interaction_2018}. Katja Kwastek also emphasizes the relevance of kinetic art in the context of interactive art development as ``Unlike Happenings and Action art, which involve the participation of the artists, Kinetic art offers opportunities for action that can be realized in the absence of the artist'' \cite[p.23]{kwastek_aesthetics_2013}.

In 1913, Marcel Duchamp produced his first kinetic participatory and ready-made \textit{BICYCLE WHEEL}. By spinning the wheel manually, the audience can experience an optical effect of the work; they witness the illusional change via the motion \cite[p.25]{weibel_it_2007}. Around the same period, between 1913 and 1920, the origins of Kinetic and OP art emerged. Peter Weibel summarises the tendencies of that time as ``the intuitive use of the algorithm concept led to mechanical and manual practices of programming, procedural instructions, interactivity, and virtuality. In the `New Tendency' shows ..., viewer participation in constructing a work of art played a considerable role'' \cite[p.25]{weibel_it_2007}. Dinkla states that participatory, reactive kinetic, and cybernetic art have much more in common with interactive art than with Happenings \cite{dinkla_participation_1996}.

\subsubsection{From Social Participation to Technological Interaction}
It is important to distinguish between participation as a social or performative gesture and interaction as technologically mediated engagement when tracing the emergence of interaction in artistic practices. In the first category, artworks based on social participation and/or the execution of instructions were common in the participatory artworks of Dada, Fluxus, and conceptual art \cite[p.4]{kwastek_aesthetics_2013}. Technological mediation has become a key feature of interactive art; artworks that use no digital technology are participatory art, and those that do are interactive art.

% \subsection{Summary}
This historical overview reveals a clear trajectory: the concept of interaction, originally rooted in sociocultural and communicative contexts, has gradually become narrowly defined through the lens of human-computer interaction. While this technological framing enabled new artistic possibilities and formalized interactive practices, it also constrained interaction within specific logics of input/output, control, and interface. This narrowing overlooks the relational, social, and affective dimensions of interaction that were present in earlier philosophical and sociological discourses. Recognizing this shift is crucial for reframing interactive art through a relational lens, which this paper proposes as a necessary move to re-engage with the fuller scope of interaction's potential.

\section{Interactive Art}\label{sec: interactiveArtEvo}

Having traced how the concept of interaction evolved from a rich sociocultural term into a narrowed technological paradigm in both scientific fields and influenced art practices, this section turns to the establishment of interactive art as a genre. While early experimental artworks engaged with cybernetic ideas and reactive systems, the formalization of interactive art brought new questions of the medium and audience engagement to the forefront. 

\subsection{Institutionalizing Interactive Art}

Although early traces of interactive practices in art date back to the 1950s, interactive art gained widespread recognition as a distinct artistic discipline in 1990, when Prix Ars Electronica introduced a dedicated competition category \cite[p.8]{kwastek_aesthetics_2013}. Roger Malina marked the moment as ``The Beginning of a New Art Form'' in that year's festival catalog \cite[p.157]{leopoldseder_prix_1990}. As the largest and oldest media art festival in the world, Ars Electronica in Linz, Austria, institutionalized Interactive Art as a genre. Myron Krueger, often credited with coining the term ``interactive art,'' was the first primary award recipient. Since then, the Prix Ars Electronica has provided an annual overview of artistic interfaces and interactive systems in the Interactive Art category, as well as in Next Idea (2004–2015) and Hybrid Arts (2007–2017). 

In 2016, the category was renamed Interactive Art+. The jury sparked that ``the language of interaction in the realm of art has matured to the extent that it can now occupy spaces outside the artists' laboratory and beyond the confines of the screen,'' and that ``interaction is the most pervasive medium that allows art to create a dialogue with every form of human activity'' \cite{de_jaeger_transcending_2016}. As Raivo Kelomees noted, audiences and juries alike became less preoccupied with whether a work was technically interactive, and conceptual and aesthetic qualities gained more weight \cite[p.53]{guljajeva_interaction_2018}.

\subsection{Crisis of Interactive Art}

Despite its institutional success, interactive art faced growing conceptual ambiguity in the 2000s. In 2004, the Golden Nica was awarded to \textit{LISTENING POST} by Mark Hansen and Ben Rubin, an artwork without any audience interaction, sparking debates around the very definition of interactivity. Erkki Huhtamo wrote about the vital phenomenon as ``Today (in 2007) interactive media is everywhere; its forms have become commonplace'' \cite{huhtamo_twintouchtestredux_2007}. Daniel Dieters mentioned his similar idea that ``Today interactivity is no longer an experiment in the media lab or an experience in a media art exhibition but part of everyday life in digital culture'' \cite[p.56]{daniels_strategies_2008}. Guljajeva continued the idea of reading it as a crisis of interactive art and further argued that audiences are ``less appreciative'' and ``much more knowledgeable'' towards interactivity (and technology) in artworks but are more interested in the concept and ``less hungry for entertainment'' \cite[pp. 11-12]{guljajeva_interaction_2018}, quoting Ryszard Kluszcznski: ``Long gone are the times of fascination just with the phenomenon of digital interactivity itself'' \cite[p.27]{kluszczynski_strategies_2010}.

When commenting on \textit{PENDULUM CHOIR} by \textit{Cod.Act} et. al, the Golden Nica award in the interactive art category in 2013, the jury stated that ``...interactivity should no longer be defined by a specific formal language prescribed by the vocabulary of the 1990s. Today, it is a set of strategies that can be situated in multiple contexts, within the art world and outside of it'' \cite[P.156]{leopoldseder_cyberarts_2014}. This aligns with the fact that Ars Electronica renamed the Interactive Art category to ``Interactive Art +'' to include a broader range of artworks in 2016. 

We share the opinion of those jury members and regard the expanding, yet increasingly undefined, scope of interactive art as a sign of its maturity rather than a crisis. Just as ``digital'' is no longer a necessary adjective in front of ``media art,'' the term ``interactive'' as human-machine or technology-mediated interaction could also fade into the background. Interaction should not only be constrained within the dualism of human versus artwork (or system and machine) but can be widely expanded to include interpersonal interaction, returning to its etymological and sociological origins.

\subsection{The Evolving Definitions}

Across the past three decades, definitions of interactive art have evolved in tandem with technological developments and audience expectations. From Myron Krueger's early emphasis on real-time human–machine systems, to Popper's insistence on audience participation as essential, and Dinkla's medium-specific framing around digital feedback systems, a consistent through line has been the active role of the audience in shaping the work. More recent interpretations recognized interactivity as a fluid set of strategies that extend across domains, formats, and intentions \cite{guljajeva_interaction_2018,leopoldseder_cyberarts_2014}. Kwastek challenges ``aesthetic distance'' and defines that ``Digital artworks that require the viewer to engage in some activity that goes beyond purely mental reception are commonly designated as `interactive art''' \cite[p.4]{kwastek_aesthetics_2013}. Dinkla defines ``interactive art'' in a medium-specific manner, as ``a genre-specific designation for computer-supported works, in which an interaction takes place between digital computer systems and users'' \cite{dinkla_pioniere_1997}. Popper states interactivity constitutes active participation, and only an artwork that involves a human-machine relation is an interactive one; the artwork is incomplete if there is no participation by the audience \cite{popper_art_1997}. These definitions claim that interaction is technology-mediated, while the artwork is responsive to audience interaction as the artist designed. The audience completes the work through a technological medium \cite[p.44]{guljajeva_interaction_2018}, which inevitably falls into the narrow view of interaction itself. 

\subsection{Redefine Interactive Art}

Rather than viewing the expanding and increasingly ambiguous scope of interactive art as a crisis, I interpret it as a sign of the field's conceptual maturity. Just as the qualifier ``digital'' has faded from media art, the term ``interactive'' no longer needs to be tethered strictly to human-machine relations. In addition, I see interactive art encompass broader relational strategies resonating with interaction's etymological and sociological roots, as elaborated in Sec. \ref{sec:cases}. Synthesizing these perspectives, I propose a working definition of interactive art as follows:

\begin{quotation}
An interactive artwork involves at least two interactive subjects and/or components, such as the audience(s) and artwork(s), multiple audiences, or even artworks themselves, engaged in a process of exchange. This interaction, typically mediated through technology, may be either bidirectional (interactive) or unidirectional (reactive), and happens in an artist-designed way, intentionally or unintentionally. Crucially, the art unfolds through this relational dynamic, with the audience or artwork playing an active role in its realization.
\end{quotation}

This definition reflects current cultural and technical conditions while remaining grounded in the historical principles of participatory engagement and system responsiveness. Rather than prescribing a rigid boundary, it acknowledges interactivity as a situated, evolving, and context-dependent phenomenon in contemporary art.

\section{Aesthetics}

The increasing emphasis on system-based interactivity and technologically mediated interaction, while advancing early interactive artwork innovations, has often overshadowed the social, affective, and intersubjective dimensions once central to interaction. This narrowing calls for a return to broader aesthetic frameworks that re-engage these lost dimensions, prompting a turn to the aesthetics of interaction and its relational potentials. Aesthetics has undergone significant shifts in the context of contemporary and interactive art, progressively expanded, from Kantian contemplation to participatory engagement, and then branching into technologically mediated interaction and socially oriented relational aesthetics. Each development brings new perspectives on how art is experienced, created, and interpreted. 

In what follows, I trace this evolution through a series of interconnected concepts, beginning with Berleant's aesthetic engagement, moving into Kwastek's aesthetics of interaction and relational aesthetics. These subsections collectively frame a broader trajectory: from contemplation to engagement, from interaction to relationality, grounding my later proposal: a return to and merging of these paths through what I term Techno Relational Aesthetics.

\subsection{Aesthetics Engagement}

The traditional aesthetic appreciative process and experience, i.e., aesthetic experience, epitomized in Kantian aesthetics, treats the aesthetic experience as ``the subjective appreciation of a beautiful object'' \cite[p.229]{berleant_art_1993}. This contemplative and distancing attitude, inherited in Kantian disinterestedness and dualism, is no longer appropriate to the emerging art and media world. Starting from the 1960s and '70s, art ``incorporated surprising subject matters'' and began to ``demand active involvement'' in its appreciative process in visual art and other art forms \cite[p.237]{berleant_art_1993}.

Within such context, philosophers and theorists shift their attention and focus from the ``art object (or artifact of the artwork)'' to the ``appreciative experience of art'' \cite{berleant_what_2013}. Arnold Berleant coined ``aesthetic engagement'' \cite{berleant_art_1993} to address such challenging developments in contemporary art. It emphasizes the ``holistic and contextual character of aesthetic appreciation'' and involves ``active participation (including physical action and creative perceptual involvement) in the appreciative process'' \cite[231]{berleant_environment_2002}. Aesthetics return to their etymological origin, primarily in the sense of perception and sensible experience across modalities such as kinesthetic and somaesthetic sensations. 

The holistic and unified aesthetics epitomize the concept of aesthetic engagement. It transcended the separations between the appreciator and the art object as well as the artist and the audience, replacing the dualism in traditional aesthetic experience. Berleant claims that aesthetic engagement ``recognizes all these functions overlap and merge within the aesthetic field,'' makes them ``become aspects of the aesthetic process'' and their ``customary separations and opposition'' disappear, which results in an ``active, direct, and intimate'' appreciative experience \cite{berleant_what_2013}. Art's beauty and aesthetic value become more general, and the relation between the appreciator and the art object (as well as the audience and artist) becomes a reciprocal process of perceptual participation.

\subsection{Aesthetics of Interaction}

Katja Kwastek describes the general usage of the term ``interaction'' as denoted ``mutual or reciprocal action or influence'' and termed it as ``reciprocity'' \cite[p.16]{kwastek_interactivity_2008}, quoting the \textit{``Dictionary of Philosophy and Psychology's''} definition of ``interaction'' as ``the relation between two or more relatively independent things or systems of change which advance, hinder, limit, or otherwise affect one another.'' \cite[p.561]{baldwin_dictionary_1901} \cite[p.16]{kwastek_interactivity_2008}. In her book \textit{Aesthetics of Interaction in Digital Art} \cite{kwastek_aesthetics_2013}, Kwastek thoroughly explored the idea of ``aesthetic of interaction'' and extends the ideas of engagement and participation into the realm of interactive media, where the relationship between the artwork and audience is dynamic and co-creative. 

Kwastek's framework acknowledges the unique ways digital art engages viewers, not just as observers but as participants who influence the outcome and form of the artwork itself. This interactivity introduces a new layer to the aesthetic experience, characterized by a loop of actions and reactions between the artwork and its audience. The aesthetics of interaction thus enrich the discourse by incorporating technological mediums and emphasizing the role of the viewer's active participation in shaping the aesthetic experience, challenging the static and passive consumption of art posited by Kantian aesthetics.

\subsection{Relational Aesthetics}

To further extend the discourse on aesthetics and interaction, Nicolas Bourriaud's concept of \textit{Relational Aesthetics} offers a significant philosophical and curatorial framework. Bourriaud proposed this notion to describe a set of artistic practices emerging in the 1990s that take human relations and their social contexts as the central material of the artwork \cite{bourriaud_relational_1998}. In contrast to traditional aesthetics concerned with formal qualities or even interactive dynamics limited to technological systems, relational aesthetics emphasizes the construction of shared spaces and social exchanges. These works prioritize encounters, participation, and conviviality over objecthood, highlighting the artwork's function as a catalyst for intersubjective experiences.

Bourriaud characterizes these practices as ``art taking as its theoretical horizon the realm of human interactions and its social context, rather than the assertion of an independent and private symbolic space'' \cite[p.14]{bourriaud_relational_1998}. This relational turn shares affinities with aesthetic engagement and interaction aesthetics, yet diverges in its explicit commitment to the social fabric of experience and its reconfiguration through artistic gestures. Artists such as Rirkrit Tiravanija, Liam Gillick, and Thomas Hirschhorn create ephemeral situations that foster social interaction, collapsing boundaries between artist, audience, and artwork. Relational aesthetics allows us to recognize practices that foreground interpersonal relations often mediated by but not reducible to technology. However, it is important to note that Bourriaud's relational aesthetics did not emerge from a technology-based art tradition. The emphasis was on ephemeral, socially engaged, and convivial situations in physical space, rather than technologically mediated interaction.

\section{Techno Relational Aesthetics}

Building upon prior aesthetic frameworks, I propose the concept of \textit{Techno Relational Aesthetics} to articulate an emerging orientation in interactive art that fuses technological mediation with interpersonal and intersubjective relationality. While Bourriaud's relational aesthetics emphasizes socially engaged practices in physical spaces, and Kwastek's aesthetics of interaction focuses on co-creative dynamics in digital media, techno-relational aesthetics draws from both yet shifts the emphasis toward emergent, hybrid configurations where human and technological agencies co-produce aesthetic experience.

This aesthetic mode foregrounds the affective and reciprocal capacities of technologically mediated interactions. In techno-relational works, technology is not merely a neutral channel but an active agent shaping the relational space. Such works often incorporate biometric sensing, adaptive AI systems, or spatial interfaces that enable feedback loops between multiple actors. The aesthetic value arises not from a pre-designed output but from the evolving, situated relational dynamic. In this sense, techno-relational aesthetics captures an emergent grammar of interactivity centered on collective perception, emotional resonance, and techno-social co-presence.

\begin{figure}[h!]
  \centering
    \begin{subfigure}{0.475\textwidth}
    \centering
    \includegraphics[width=\textwidth,keepaspectratio]{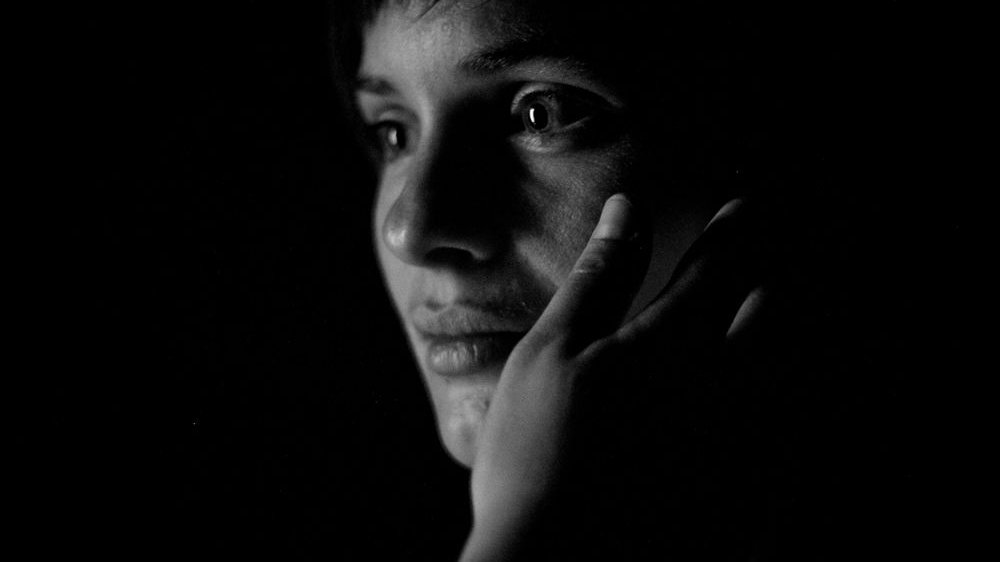}
      \caption{\textit{SENSITIVE TO PLEASURE}, 2011. \copyright{Sonia Cillari}.}
      \label{fig: SENSITIVE}
    \end{subfigure}
    \hfill
    \begin{subfigure}{0.475\textwidth}
    \centering
      \includegraphics[width=\textwidth,keepaspectratio]{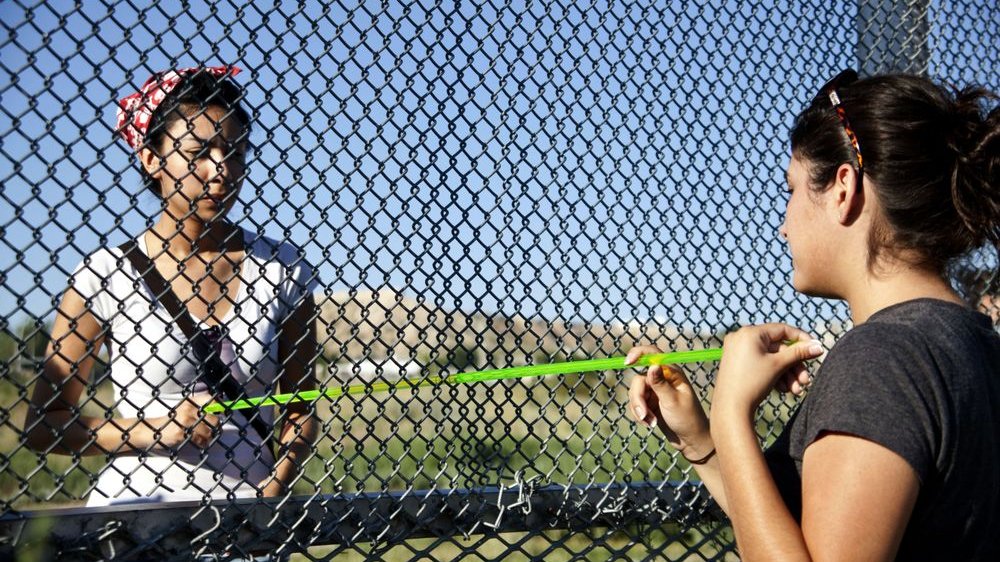}
      \caption{\textit{CROSS COORDINATES (MX-US)}, 2012. \copyright{Iván Abreu}.}
      \label{fig: CROSS}
    \end{subfigure}
        \hfill
    \begin{subfigure}{0.475\textwidth}
    \centering
    \includegraphics[width=\textwidth,keepaspectratio]{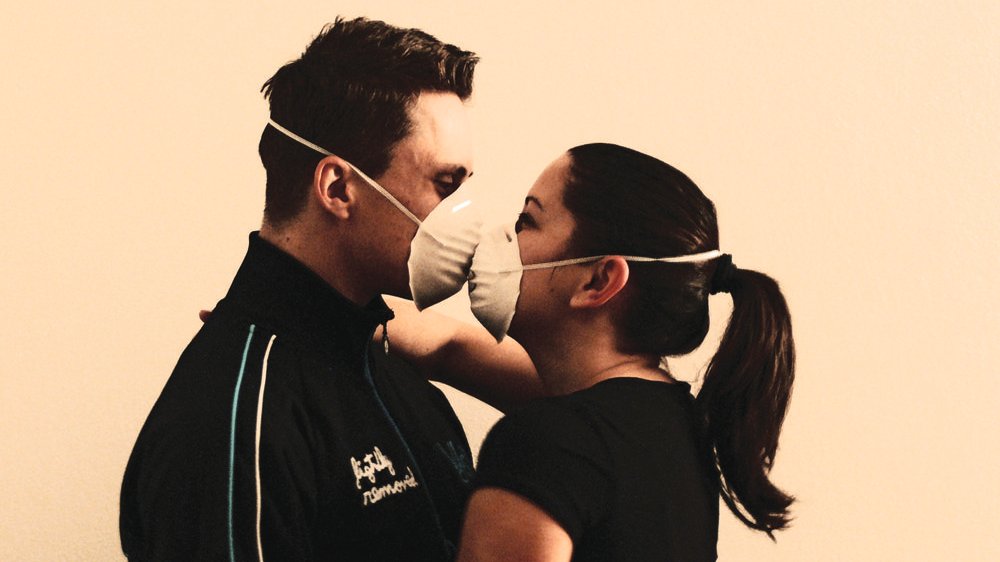}
      \caption{\textit{FUTURE KISS}, 2009. \copyright{Lenka Klimesova}.}
      \label{fig: KISS}
    \end{subfigure}
        \hfill
  \begin{subfigure}{0.45\textwidth}
        \centering
      \includegraphics[width=\textwidth,keepaspectratio]{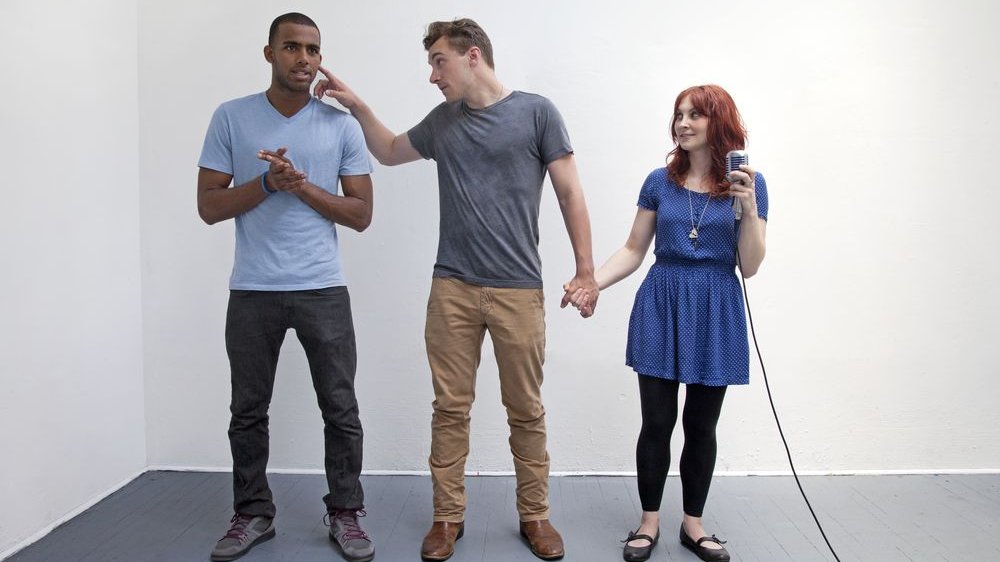}
      \caption{\textit{ISHIN-DEN-SHIN}, 2013. \copyright{Olivier Bau et al.}}
      \label{fig: ISHIN}
  \end{subfigure}
%     \caption{Direct-Interpersonal Interaction (DII) (Duo): Artworks involving real-time interaction among audience members in pairs.}
%     \label{fig: indirect}
% \end{figure}
% \begin{figure}[h!]
%   \centering
    \begin{subfigure}{0.475\textwidth}
    \centering
    \includegraphics[width=\textwidth,keepaspectratio]{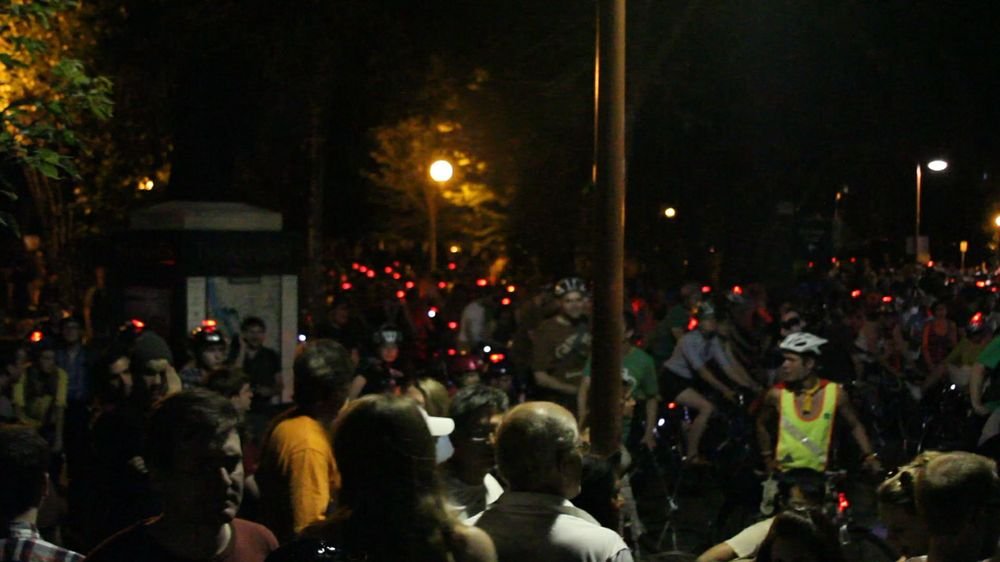}
      \caption{\textit{THE KURAMOTO MODEL}, 2013. \copyright{David Allan Rueter}.}
      \label{fig: MODEL}
    \end{subfigure}
    \hfill
    \begin{subfigure}{0.475\textwidth}
    \centering
      \includegraphics[width=\textwidth,keepaspectratio]{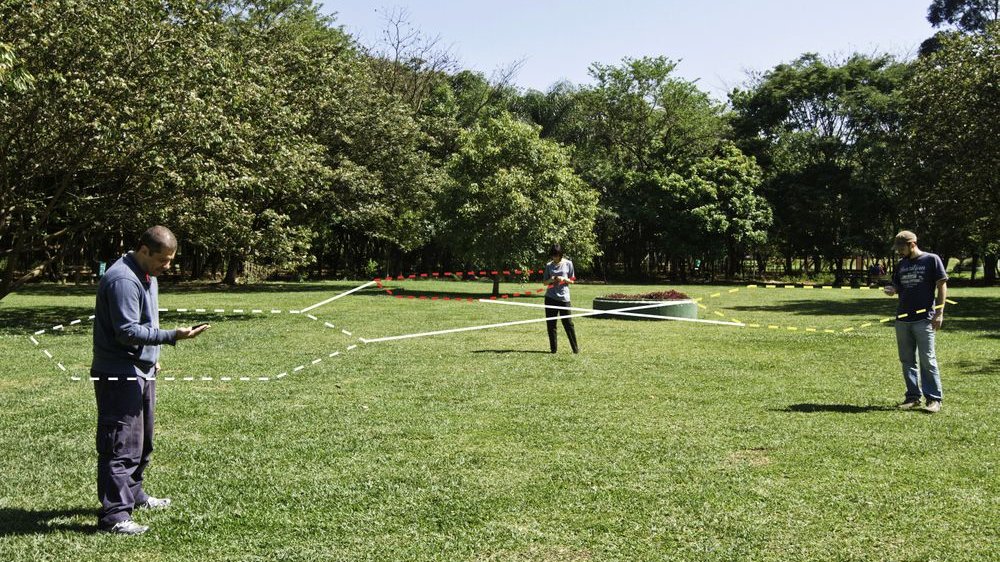}
      \caption{\textit{REDES VESTÍVEIS/WEARABLE NETS}, 2011. \copyright{Claudio Bueno}.}
      \label{fig: WEARABLE}
    \end{subfigure}
    \hfill
    \begin{subfigure}{0.5\textwidth}
    \centering
    \includegraphics[width=\textwidth]{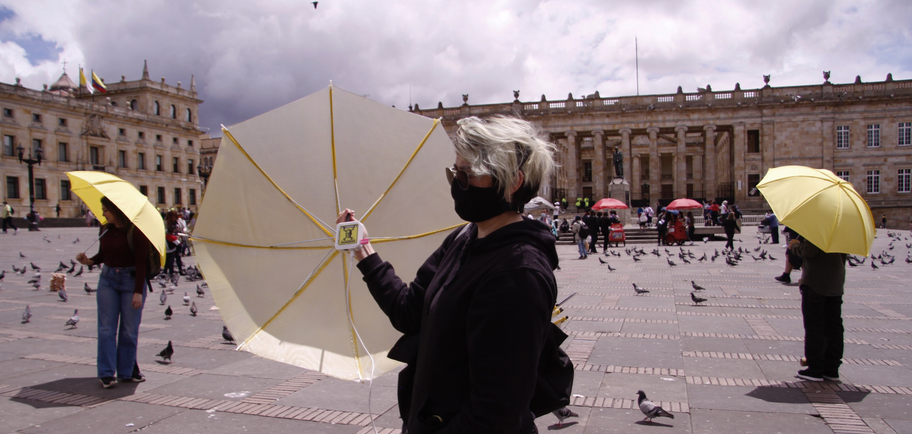}
      \caption{\textit{BI0FILM.NET: RESIST LIKE BACTERIA}, 2022. \copyright{Hsu and Rivera}.}
      \label{fig: biofilm}
    \end{subfigure}
    \caption{Various Interpersonal Interaction Among Audience Members.}
    % \description{Various Interpersonal Interaction Among Audience Members.}
    \label{fig: direct}
\end{figure}

\section{Case Studies from Ars Electronica} \label{sec:cases}

To ground the theoretical propositions of techno relational aesthetics, this section draws on selected case studies from Ars Electronica, a pivotal platform for showcasing experimental and interactive media art. These examples demonstrate how interaction is variously framed—as bodily resonance, spatial responsiveness, social mediation, and intersubjective exchange—often blurring the boundary between audience and system, between artwork and relational event.

Several artworks extend spatial and bodily interactivity to multiple audience members. In \textit{AHORA} by Kerllenevich and Alsina, audience movement through space modulates a layered auditory environment, fostering personalized soundscapes. Similarly, Cantoni and Crescenti's \textit{TUNNEL} uses physical pressure from viewers to set kinetic ripples in motion across portico structures, making spatial interaction visible and affectively charged. Lucy McRae's \textit{COMPRESSION CRADLE} explores intimacy and containment, simulating human embrace via aerated cushions that contour to the participant's body. In Bambozzi's \textit{MOBILE CRASH}, physical gestures—pointing to screens—trigger cinematic sequences, establishing direct spatial agency. Social and performative dynamics surface strongly in works like \textit{PENDULUM CHOIR} by Cod.Act, where a group of singers is suspended on hydraulic platforms, creating a choreography between vocal output and mechanical motion. \textit{INFERNO} by Demers and Vorn deepens this tension between autonomy and control, as audience members wear robotic exoskeletons that choreograph their movements during a live performance. These works situate interactivity in the negotiation between bodies, systems, and social scripts.

Other works emphasize temporally situated or affectively collective interaction through techno-interpersonal engagements. In \textit{MIND} by Shinseungback Kimyonghun, facial expressions of recent participants shape generative audio reminiscent of ocean waves, constructing a collective mood archive. Rafael Lozano-Hemmer's \textit{VOICE ARRAY} stores participants' voices in a visual light-based memory structure, where each new utterance pushes older ones along a sequence of illuminated nodes—producing a visible, accretive form of audience memory. Some artworks facilitate inter-audience connection. \textit{ECHO} by Pinn and Rossi lets participants choose another's face in a virtual mirror to hear their personal narrative, literally facing the stories of others. In \textit{FUTURE KISS}, vibrating masks with kiss detectors invite participants to find and kiss their ``other half.'' \textit{ISHIN-DEN-SHIN} transmits whispered audio via the human body, turning physical touch into a conduit for intimate sonic exchange.

At the level of distributed collectivity and interpersonal dynamics, \textit{THE KURAMOTO MODEL (1,000 FIREFLIES)} synchronizes the blinking of thousands of radio-connected bike lights across a city, generating emergent visual harmony. \textit{REDES VESTÍVEIS / WEARABLE NETS} by Claudio Bueno invites participants into a virtual net visualized via smartphones, emphasizing collective spatial coordination. Hsu and Rivera's \textit{BI0FILM.NET} transforms protest umbrellas into decentralized WiFi antennas—using interaction as a tool for resistance and communication infrastructure. Together, these works point towards the relational turn and anchor the paper's theoretical claims in situated practices that redefine interaction not as input–output exchange but as relationality.

\section{A Relational (Re)Turn}
\label{sec: inflation_of_interactivity}

As discussed earlier, interaction has narrowed from its original sociocultural meanings, i.e., reciprocity, mutual influence, and intersubjective dynamics, toward a technocentric paradigm dominated by human-computer interaction. Meanwhile, aesthetic discourse has expanded from Kantian contemplation to participatory engagement, branching into techno-centered interactivity and socially-oriented relational aesthetics. These trajectories open a space for reconceptualizing interaction beyond technical mechanisms, but as situated, affective, and co-constitutive experience.

\subsection{Relational Interaction}

We call for intersubjective and interpersonal dynamics as the etymological and relational return to interaction and interactive art. As previously discussed in Sec. \ref{sec: interaction}, the concept of interaction originated in sociocultural and communicative contexts, highlighting mutual influence, shared experience, and symbolic exchange. Over time, however, the notion of interaction was progressively narrowed within the human-computer and audience-artwork interaction paradigm, emphasizing input and output dynamics, interface control, and system responsiveness. While this technological framing enabled the emergence of interactive art as a formal genre, it also narrowly constrained interaction to mechanical.

By contrast, relational interaction reclaims the original breadth of the term. It prioritizes the formation of shared meaning, co-presence, and reciprocal engagement among participants, often mediated through, but not reducible to, artworks or technological systems. In this new framing, interaction becomes an embodied and intersubjective process of co-existence and mutual recognition. It can manifest in various artwork-mediated configurations, particularly in audience-audience engagement, producing dynamic interpersonal engagements where the artwork serves as an interface of relational exchange rather than a passive display system.

Relational interaction reframes interactivity as a socially situated and affectively resonant phenomenon by shifting focus away from isolated audience-artwork feedback loops and toward collective dynamics. It re-imagines interaction beyond manipulation of interactivity as a shared aesthetic and communicative encounter, recalling its etymological and theoretical roots while opening new possibilities for interactive art in contemporary contexts.

\subsection{Towards Expanded Relationality}

Building on the re-emphasis of relational interaction in interpersonal terms, this subsection further proposes an expanded understanding of relationality that moves beyond interpersonal and human–human relations toward a broader spectrum of intersubjective configurations, including interspecific, more-than-human, and planetary-scale entanglements. The redefinition of what counts as a ``subject'' in interaction opens the aesthetic and conceptual framework of interactive art to include nonhuman agencies, ecological systems, and distributed materialities. In this expanded frame, interaction becomes a site of encounter between heterogeneous entities, grounded in intersubjective responsiveness rather than control. 

Philosophers such as Donna Haraway and Karen Barad have emphasized how intersubjectivity must be reconceived as an entangled intra-action, where entities do not precede their relations but rather emerge through them. This shift resonates with contemporary interactive artworks that involve plants, animals, microbial agents, AI systems, or data ecologies as co-creators or participants in aesthetic processes. This relational (re)turn toward expanded relationality extends the aesthetics and ethics of interaction, positioning interaction as a more-than-human practice, where affect, agency, and co-presence are distributed across human and nonhuman actors alike.

\section{Conclusion} \label{sec: discuss}

This paper proposes a relational (re)turn and techno relational aesthetics in interactive art that working towards expanded relationality in art practice and theory. It addresses the limitations of current human-computer paradigms by reconfiguring interaction through the lens of intersubjectivity in interaction and relationality in aesthetics. The paper critiques the narrowing of ``interaction'' into HCI-based frameworks and previous interactive art studies, which privilege input/output, control, and responsiveness. It argues for a reclamation of interaction's sociocultural roots, emphasizing mutual influence, reciprocity, and symbolic exchange. This reorientation centers interaction as a relational and communicative encounter, not a technical function. The evolution of aesthetic theory culminates in the paper's proposal of Techno Relational Aesthetics, which synthesizes technological mediation as co-productive of affective, social, and intersubjective experiences. Rather than limiting interactivity to audience–artwork feedback loops, the paper foregrounds artwork-mediated audience–audience interaction in interactive art, where interaction becomes a social, embodied process and the artwork serves as the relational mediation. The final claim of this paper extends relationality beyond the interpersonal and introduces expanded relationality as a framework toward interspecific, ecological, and planetary-scale interactions. Thus, interaction and interactive art can continue the practice of intersubjectivity and include nonhuman actors (plants, AI, data systems), leading to a co-becoming of more-than-human entanglements.

%% The next two lines define the bibliography style to be used and
%% the bibliography file.
\bibliographystyle{ACM-Reference-Format}
\bibliography{sample-base}

\end{document}